\documentclass[showpacs,twocolumn,aps,preprintnumbers,letterpaper]{revtex4}
\usepackage{amsmath,amssymb}
\usepackage{epsfig}
\usepackage{graphicx}
\usepackage{amsmath}
\usepackage{slashed}
\usepackage{amsfonts}
\usepackage{epstopdf}

\newcommand{\be}{\begin{equation}}
\newcommand{\ee}{\end{equation}}
\newcommand{\bear}{\begin{eqnarray}}
\newcommand{\eear}{\end{eqnarray}}
\newcommand{\ba}{\begin{array}}
\newcommand{\ea}{\end{array}}

\def\be{\begin{eqnarray}}
\def\ee{\end{eqnarray}}
\def\bea{\be}
\def\eea{\ee}

\def\roughly#1{\mathrel{\raise.3ex\hbox{$#1$\kern-.75em%
\lower1ex\hbox{$\sim$}}}}

\begin{document}

\title{Holographic Heavy-Light Chiral Effective Action}

\author{Yizhuang Liu}
\email{yizhuang.liu@stonybrook.edu}
\affiliation{Department of Physics and Astronomy, Stony Brook University, Stony Brook, New York 11794-3800, USA}

\author{Ismail Zahed}
\email{ismail.zahed@stonybrook.edu}
\affiliation{Department of Physics and Astronomy, Stony Brook University, Stony Brook, New York 11794-3800, USA}


\date{\today}
\begin{abstract}
We propose a  variant of the  $D4$-$D8$ construction to describe  the low energy effective theory of heavy-light mesons, interacting
with the lowest lying pseudoscalar and vector mesons. The  heavy degrees of freedom are identified with the 
$D8_L$-$D8_H$ string low energy modes, and are approximated near the world volume of  $N_f-1$
light $D8_L$ branes, by fundamental vector field valued in $U(N_f-1)$. The effective  action follows from the reduction of 
the  bulk Dirac-Born-Infeld (DBI) and  Chern-Simons (CS) actions, and is shown to exhibit both chiral and heavy-quark symmetry. The action interpolates
continuously between the $U(N_f)$ case with massless mesons,  and the $U(N_f-1)$ case with heavy-light mesons. 
The heavy-light meson radial spectrum is Regge-like.
The one-pion and two-pion couplings  to the heavy-light multiplets 
are evaluated. The partial widths for the charged decays
$G\rightarrow H+\pi$ are shown  to be comparable to the recently  reported  full widths
for both the charm and bottom mesons. 
 \end{abstract}


\maketitle

\setcounter{footnote}{0}


\section{Introduction}
In QCD the light quark sector (u, d, s) is dominated by the spontaneous breaking of chiral
symmetry. The heavy quark sector (c, b, t) is characterized by heavy-quark symmetry~\cite{ISGUR}. 
The combination of both symmetries led to the conclusion that the heavy-light
doublet $(0^-, 1^-)=(D,D^*)$ has a chiral partner $(0^+, 1^+)=(\tilde D,{\tilde D}^*)$ that is
about  one consituent mass heavier~\cite{MACIEK,BARDEEN}. This observation 
is supported by  the  BaBar collaboration~\cite{BABAR}  and the CLEOII collaboration~\cite{CLEOII}.

More recently,
the Belle collaboration~\cite{BELLE} and the BESIII collaboration~\cite{BESIII} have reported the observations
of multiquark exotics,  with quantum numbers uncommensurate with the excited states of charmonia and bottomia, such as 
the neutral $X(3872)$ and the charged $Z_c(3900)^\pm$ and $Z_b(10610)^\pm$ to cite afew.  These sightings and more
have been supported by the DO collaboration at Fermilab~\cite{DO},  and the  LHCb collaboration at CERN~\cite{LHCb}.
They provide a  window to  new phenomena involving heavy-light multiquark states.

Theoretical arguments have predicted the 
occurence of some of these exotics as molecular bound states mediated by one-pion exchange much
like deuterons or deusons~\cite{MOLECULES,THORSSON,KARLINER,OTHERS,OTHERSX,OTHERSZ,OTHERSXX,LIUMOLECULE}.
Non-molecular heavy exotics were also discussed using constituent quark models~\cite{MANOHAR}, 
heavy solitonic baryons~\cite{RISKA,MACIEK2}, instantons~\cite{MACIEK3} and QCD sum rules~\cite{SUHONG}. The molecular
mechanism favors the formation of shallow bound states near treshold, while the non-molecular
or quarkonium mechanism leads to deeply bound states.

The holographic approach offers a useful framework
for discussing both the spontaneous breaking of chiral symmetry and confinement, in the 
double limit of large $N_c$ and large t$^\prime$Hooft coupling $\lambda=g^2N_c$.
An example is the $D4$-$D8$ model  suggested by Sakai and Sugimoto~\cite{SSX}.
In short, the model consists of $N_f$ probe $D8$ and $D\bar 8$ branes  in a background of $N_c$ $D4$ branes. 
The induced gravity on the probe branes, cause them to fuse in the infrared providing a geometrical mechanism for
the spontaneous breaking of chiral symmetry. The DBI
 action on the probe branes, provides a low-energy effective
action for the light pseudoscalars with full global chiral symmetry, where the vectors and axial-vector light mesons 
are dynamical gauge particles of a hidden chiral symmetry~\cite{HIDDEN}.

The purpose of this paper is to address the dual concepts of chiral and heavy-quark symmetries by
using the holographic construction. We will show that a variant of the $D4$-$D8$ construct,  composed of
$(N_f-1)$ light and one heavy probe branes allows a geometrical set up for the derivation of the
leading heavy-light (HL)  effective action in conformity with chiral and heavy-quark symmetry. The heavy-light
mesons are identified with the string low energy modes, and approximated
by bi-fundamental and local  vector fields in the vicinity of the light probe branes. Their masses follow from the vev of the
moduli span by the dilaton fields in the DBI action. We note that few approaches were proposed for the 
description of heavy-light mesons using holography without the strictures of chiral symmetry~\cite{FEWX,BRODSKY}.

The organization of the paper is as follows: In section 2 we 
briefly outline  the geometrical set up  for the derivation of the heavy-light  effective action (HL).
We identify the pertinent light and heavy fields and explicit their contributions to the (expanded)
DBI and CS actions. 
In section 3, we detail the analysis of the HL meson spectrum. 
In section 4, we derive the chiral interactions to the HL mesons and deduce their 
corresponding  axial couplings within and across the HL meson multiplets. We use these couplings to estimate
the HL charm and bottom one pion charged decays. 
Our conclusions are in section 5. 

\section{ Holographic effective action}

\subsection{D-brane set up}

\begin{figure}[h!]
\begin{center}
 \includegraphics[width=6cm]{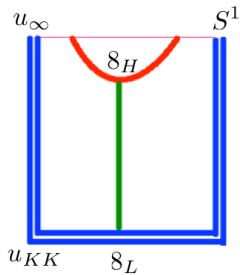}
  \caption{$N_f-1=2$ antipodal $8_L$ light branes, and 1 $8_H$  heavy brane shown in the $\tau U$ plane, with a massive
$HL$-string connecting them. When the latter is massless, the $8_H$ brane coincides with the $8_L$ branes, transmuting
  to $N_f$ $8+\bar 8$ light branes.}
  \label{fig_branex}
 \end{center}
\end{figure}

The $D4$-$D8$ construction proposed by Sakai and Sugimoto~\cite{SSX}
for the description of the spontaneous breaking of chiral symmetry and the
ensuing chiral effective action is by now well-known, and will not be repeated here. Instead, 
consider the variant with  $N_f-1$ light $D8$-$\bar D8$ (L) and one heavy (H) probe branes in the cigar-shaped geometry that
spontaneously breaks chiral symmetry. Also and for simplicity, the light probe branes are always assumed
in the anti-podal configuration. A schematic description  of the set up for $N_f=3$ is shown in Fig.~\ref{fig_branex}.
We assume that the L-brane world volume consists of $R^{4}\times S^1\times S^4$ with 
$[0-9]$-dimensions.  The light 8-branes are embedded in the $[0-3+5-9]$-dimensions and set
at the antipodes of $S^1$ which lies in the 4-dimension. 
The warped $[5-9]$-space is characterized by a finite size  $R$ and a horizon at $U_{KK}$.

\subsection{DBI and CS actions}

The lowest open string modes streched between the H- and L-branes
as shown in Fig.~\ref{fig_branex}, when viewed near the L brane world volume, consist of tranverse modes $\Phi_M$ and longitudinal modes $\Psi$, both fundamental with respect to the flavor group $SU(N_f-1)$. 
At non-zero brane separation, these fields acquire  a vev that makes the vector field massive~\cite{MEYERS}.  Strictly speaking these
fields are bi-local, but near the L-branes we will approximate them  by local vector fields that are described by the standard DBI action. 
In this respect, our construction is distinct from the approaches developed  in~\cite{FEWX}.

With this in mind and to leading order in the $1/\lambda$ expansion, the effective action on the probe L-branes
consists of the non-Abelian DBI  (D-brane Born-Infeld) and CS (Chern-Simons) action.  After integrating over the $S^4$, the leading contribution to the DBI action is

\bea
\label{1}
S_{\rm DBI}\approx -\kappa\int d^4x dz\,{\rm Tr}\left(f(z){\bf F}_{\mu\nu}{\bf F}^{\mu\nu}+g(z){\bf F}_{\mu z}{\bf F}^{\nu z}\right)
\eea
The warping factors are 

\be
f(z)=\frac{R^3}{4U_z}\,,\qquad g(z)=\frac{9}{8}\frac{U_z^3}{U_{KK}}
\ee 
with $U_z^3=U_{KK}^3+U_{KK}z^2$, and $\kappa\equiv \tilde T(2\pi\alpha^\prime)^2$~\cite{SSX}.
The effective fields in the field strengths are
($M,N$ run over $(\mu,z)$)

\bea
\label{2}
&&{\bf F}_{MN}=\nonumber \\ 
&&\left(\begin{array}{cc}
F_{MN}-\Phi_{[M}\Phi_{N]}^{\dagger}&\partial_{[M}\Phi_{N]}+A_{[M}\Phi_{N]}\\
-\partial_{[M}\Phi^{\dagger}_{N]}-\Phi^{\dagger}_{[M}A_{N]}&-\Phi^{\dagger}_{[M}\Phi_{N]}
\end{array}\right)
\eea
Specifically, using (\ref{2})) we can recast the trace contribution in (\ref{1}) in the form
$f{\cal L}_f+g{\cal L}_g$ with

\bea
\label{2XX2}
&&{\cal L}_f={\rm Tr}(F_{\mu\nu}-a^{\mu\nu})(F^{\mu\nu}-a^{\mu\nu})-2f_{\mu\nu}^{\dagger}f^{\mu\nu}+b^{\mu\nu}b_{\mu\nu}\nonumber\\
&&{\cal L}_g={\rm Tr}(F_{\mu z}-a_{\mu z})(F^{\mu z}-a^{\mu z})-2f_{\mu z}^{\dagger}f^{\mu z}+b^{\mu z}b_{\mu z}\nonumber\\
\eea
and

\bea
\label{2XX3}
&&a_{MN}=\Phi_{[M}\Phi^{\dagger}_{N]}\nonumber\\
&&f_{MN}=\partial_{[M}\Phi_{N]}+A_{[M}\Phi_{N]}\nonumber\\
&&b_{MN}=\Phi^{\dagger}_{[M}\Phi_{N]}
\eea
The CS contribution to the effective action is (form notation used)

\be
\label{6}
S_{\rm CS}=\frac{N_c}{24\pi^2}\int_{R^{4+1}}{\rm Tr}\left({\bf A}{\bf F}^2-\frac{1}{2}{\bf A}^3{\bf F}+\frac{1}{10}{\bf A}^5\right)
\ee
where the normalization to $N_c$ is fixed by integrating the $F_4$ RR flux over the $S^4$.
The  matrix valued 1-form gauge field is
\be
\label{7}
{\bf A}=\left(\begin{array}{cc}
A&\Phi\\
-\Phi^{\dagger}&0
\end{array}\right)
\ee


For $N_f$ coincidental branes, the $\Phi$ multiplet is massless. However, their brane world-volume 
supports an adjoint and traceless scalar $\Psi$ in addition to the adjoint gauge field $A_M$ both of which are
hermitean and $N_f\times N_f$ valued, which we have omitted from the DBI action  in so far
for simplicity. Their leading contribution from the DBI action  is of the form (omitting the warping factors)

\be
\label{8X1}
\frac 12 {\rm Tr}\left|\nabla_M\Psi\right|^2 -\frac 14 {\rm Tr}\left([\Psi,\Psi]^2\right)
\ee
with $\nabla_M\Psi=\partial_M\Psi+i[A_M,\Psi]$. The extrema of the potential contribution in (\ref{8X1}) 
or $[[\Psi,\Psi], \Psi]=0$ define a moduli~\cite{MEYERS}.  For $N_f-1$ light branes separated from one
heavy brane, we identify one of the moduli solution with a finite vev $v$ as,

\bea
\label{8X2}
\Psi=\left(\begin{array}{cc}
-\frac v{N_f-1}{\bf 1}_{N_f-1}&0\\
0&v \end{array}\right)
\eea
Since the upper block diagonal contribution commutes with $A_\mu$, only the 
$\Phi$ multiplet acquires a Higgs-like mass through the first contribution 
in (\ref{8X1})

\bea
\label{8X3}
\frac 12 M^2 {\rm Tr}\left(\Phi^\dagger_M \Phi_M\right)\equiv \frac {v^2N_f^3}{(N_f-1)^2}{\rm Tr}\left(\Phi^\dagger_M \Phi_M\right)
\eea
The vev is related to the separation between the light and heavy branes~\cite{MEYERS}, 
which we take it to be the length of of the HL string of mass $M$, i.e. $v\sim M$. 
Below, $M$ is also the  degenerate (in practice mean) heavy meson
mass for the bi-fundamental field $\Phi$ which will be identified with the HL meson doublets,
such as   $(D, D^*)$ or  $(B, B^*)$. Throughout, we will refer to the large mass limit also as 
the heavy quark limit.

\section{HL meson spectrum}

To investigate the holographic spectrum of the heavy-light mesons, we first specialize to the case 
where the simplified  gauge potential 1-form (\ref{7})

\be
\label{7}
{\bf A}\rightarrow \left(\begin{array}{cc}
0&\Phi\\
-\Phi^{\dagger}&0
\end{array}\right)
\ee
is inserted in the unexpanded DBI action

\be
\label{7Y}
&&-{\kappa}\int d^4xdz\sqrt{-g}\nonumber\\
&&\times \left({\rm det}\left(g_{MN}+2\pi\alpha^\prime F_{MN}+\nabla_M\Psi^\dagger\nabla_N\Psi\right)\right)^{\frac 12}
\ee
As we noted earlier, the warped mass term in (\ref{8X3}) follows by using (\ref{8X2}) and expanding (\ref{7Y}) in leading order

\be
\label{7X7}
-\tilde\nu^2\int d^4x dz (U_z^2)(g^{zz}\Phi^{\dagger}_z\Phi_z+g^{xx}\Phi^{\dagger \mu}\Phi_{\mu})
\ee
Here we have set $\tilde\nu^2=\tilde T\nu^2$ with the HL mass parameter $\nu\sim v$, and 
defined the warpings as

\bea
&&U_z^2g^{zz}=\frac{9}{4}\left(\frac{U_z}{R}\right)^{\frac{3}{2}}\frac{U_z^3}{U_{KK}}\equiv 
\frac{9}{4}a(z)\frac{U_z^3}{U_{KK}}\nonumber\\
&&U_z^2g^{xx}=U_z^2\left(\frac{R}{U_z}\right)^{\frac{3}{2}}=\left(\frac{U_z}{R}\right)^{\frac{3}{2}}\frac{R^3}{U_z}
\eea
The corresponding field-strength 2-form (\ref{2}) is

\bea
\label{22X}
{\bf F}_{MN}\rightarrow \left(\begin{array}{cc}
-\Phi_{[M}\Phi_{N]}^{\dagger}&\partial_{[M}\Phi_{N]}\\
-\partial_{[M}\Phi^{\dagger}_{N]}&-\Phi^{\dagger}_{[M}\Phi_{N]}
\end{array}\right)
\eea
Inserting (\ref{22X}) into (\ref{1}) and combining it with (\ref{7X7}) yield to quadratic order

\bea
\label{22X1}
&&\frac{S_\Phi}{2\kappa}=\nonumber\\
&&-\int dz d^4x f(z)(\partial_{\mu}\Phi^{\dagger}_{\nu}-\partial_{\nu}\Phi^{\dagger}_{\mu})(\partial^{\mu}\Phi^{\nu}-\partial^{\nu}\Phi^{\mu})\nonumber \\
&&-\int dz d^4x g(z)(\partial_{\mu}\Phi^{\dagger}-\partial_{z}\Phi^{\dagger}_{\mu})(\partial^{\mu}\Phi-\partial^{z}\Phi^{\mu})\nonumber \\
&&-\tilde \nu^2\int dz d^4xa(z)(2f(z)\Phi^{\dagger \mu}\Phi_{\mu}+g(z)\Phi^{\dagger}\Phi)
\eea

\subsection{Mode analysis}

To find the HL mass spectrum, we first make  use of the general decomposition 

\be
\Phi_{\mu}=\epsilon_{\mu}(z)e^{ip\cdot x},\qquad \Phi=-i\epsilon(z)e^{ip\cdot x}
\ee
into the equations of motion following from (\ref{22X1}) to obtain

\bea
\label{22X3}
&&2f(p^2\epsilon^{\mu}-p^{\mu}\epsilon \cdot p)
+\frac{d}{dz}\left(g\left(p^{\mu}\epsilon-\frac{d\epsilon^{\mu}}{dz}\right)\right)+2af\epsilon^{\mu}=0
\nonumber\\
&&g\left(\epsilon p^2-p\cdot \frac{d\epsilon}{dz} \right)+ag\epsilon=0
\eea
We can simplify (\ref{22X3}) by redefining

\bea
\label{22X4}
&&\frac{d\tilde \epsilon_{\mu}}{dz}=\frac{d\epsilon_{\mu}}{dz}-p_{\mu}\epsilon\nonumber\\
&&\epsilon_{\mu}=\tilde \epsilon_{\mu}+p_{\mu}\int dz \epsilon
\eea
in terms of which (\ref{22X3}) now reads

\bea
\label{22X5}
&&-\frac{d}{dz}\left(g\frac{d\tilde \epsilon_{\mu}}{dz}\right)\nonumber\\
&&+2f(p^2\tilde \epsilon_{\mu}-p^{\mu}(p \cdot \tilde \epsilon))+2\tilde \nu^2af\tilde \epsilon_{\mu}+2a\tilde \nu^2fp_{\mu}\int dz \epsilon=0\nonumber\\
&&-gp\cdot\frac{d\tilde \epsilon}{dz}+a\tilde \nu^2g\epsilon=0
\eea

\subsubsection{Transverse modes}

The transverse modes solution to (\ref{22X5}) with $\tilde \epsilon \cdot p=0$,  yields $\epsilon=0$ and 
$\tilde\epsilon$ satisfying 

\be
\label{22X6}
-\frac{d}{dz}\left(g\frac{d\tilde \epsilon_{\mu}}{dz}\right)+2f(p^2+\tilde \nu^2a)\tilde \epsilon_{\mu}=0
\ee
for $\tilde \nu,a\neq 0$. The heavy-light states correspond to the normalizable bulk modes with
$p^2=-m_{n}^2$. Let $\phi_{n}(z)$  denote these normalizable modes. They satisfy the warped and massive
eigenvalue equation

\be
\label{22X7}
-\frac{d}{dz}\left(g\frac{d \phi_{n}}{dz}\right)+2f(-m_{n}^2+\tilde \nu^2a)\phi_{n}=0
\ee
We note that for coincidental branes with zero vev or $\tilde\nu=0$, 
(\ref{22X7}) reduces to the equation for the pionic zero mode in the 
massless limit  with $m_n=0$. As a result, the spontaneous breaking
of chiral symmetry will be enlarged from $SU(N_f-1)$ to $SU(N_f)$,
in the limit of $N_f$ coincidental branes in agreement with the original
analysis in~\cite{SSX}. This point will be emphasized further below
in the ensuing chiral effective action.

In terms of the modes (\ref{22X7}), the transverse mode decomposition is

\bea
\label{22X77}
&&\Phi_{\mu}(x,z)=\sum_{n=1}\phi_{n}(z)B^{n}_{\mu}(x)\nonumber\\
&&\Phi_z(x,z)=0
\eea
Inserting (\ref{22X77}) into  the vector contribution in (\ref{22X1}) we obtain

\bea
\label{22X78}
&&2\kappa\sum_{n,m}\int dz f\phi_{m}\phi_{n}\int d^4x (\partial_{[\mu}B^{m}_{\nu]})^{\dagger}(\partial^{[\mu}B^{n,\nu]})\nonumber \\
&&+2\kappa\sum_{n,m}\int(2a\tilde \nu^2f\phi_{m}\phi_n+g\phi^{\prime}_n\phi^{\prime}_m) dz \int d^4x(B^{m}_{\mu})^{\dagger}B^{n,\mu}\nonumber \\
\eea
We now note that the coefficient of the second contribution in (\ref{22X78}) can be integrated by parts to satisfy the identity

\be
\int dz\, (2a\tilde \nu^2f\phi_n-\frac{d}{dz}(g\frac{d\phi_n}{dz}))\phi_m=2m_n^2\int dz f \phi_m\phi_n
\ee
thanks to (\ref{22X4}). This  suggests to normalize $\phi_n$ as

\be
4\kappa\int dz f\phi_m\phi_n=\delta_{m,n}
\ee
which brings the quadratic HL vector contribution (\ref{22X78}) to the canonical form

\be
\sum_n\int d^4x
\left(\frac{1}{2}\partial_{[\mu}B^{n\dagger}_{\nu]}\partial^{[\mu}B^{n\nu]}+m_n^2B^{n\dagger}_{\mu}B^{n\mu}\right)
\ee

\subsubsection{Longitudinal modes}

The longitudinal modes correspond to $\tilde \epsilon \cdot p\neq 0$. For 
$p^2=-m^2$ nonzero, these modes are of the form $\tilde \epsilon^{\mu} =p^{\mu}\epsilon_1$.
In this case both $\epsilon, \epsilon_1$ are non-zero and satisfy the coupled equations

\bea
\label{22X8}
&&-\frac{d}{dz}\left(g\frac{d\epsilon_1}{dz}\right)+2\tilde \nu^2af \epsilon_1+2a\tilde \nu^2f\int \epsilon=0\nonumber\\
&&a\tilde \nu^2\epsilon+m^2\frac{d\epsilon_1}{dz}=0
\eea
from which we have 

\be
\label{22X9}
\epsilon=-\frac{d\epsilon_1}{dz}+\frac{d}{dz}\left(\frac{1}{2a\tilde \nu^2f}\frac{d}{dz}\left(g\frac{d\epsilon_1}{dz}\right)\right)
\ee
or equivalently

\be
\label{22X10}
m^2\epsilon=a\tilde \nu^2\epsilon-\frac{d}{dz}\left(\frac{1}{2af}\frac{d}{dz}(ag\epsilon)\right)
\ee
We note that (\ref{22X10}) does not lead to the pionic zero mode in bulk for
$\tilde\nu, m\rightarrow 0$, in  contrast to (\ref{22X7}). This is consistent with
the counting of Golstone modes in the limit when all the D-branes are coincidental.

In terms of the original fields $\Phi_{\mu}$ and $\Phi_{z}$, the expansion now reads

\bea
\label{22X100}
&&\Phi_z(x,z)=\sum_{n}\epsilon_n(z)D_{n}(x)\nonumber\\
&&\Phi_{\mu}(x,z)=\sum_{n}\frac{-1}{2afm_n^2}\frac{d}{dz}(ag\epsilon_n)\partial_{\mu}D_n(x)
\eea
after using  the relation between $\tilde \epsilon$ and $\epsilon$ in (\ref{22X5}).
Inserting (\ref{22X100}) into the pertinent quadratic parts in (\ref{22X1}) yields

\bea
\label{22X111}
&&2\kappa\sum_{m,n}\int dz\,ag \frac{\tilde \nu^2\epsilon_m\epsilon_n}{m_n^2}
\int d^4x \partial_{\mu}D_n^{\dagger}\partial^{\mu}D^m\nonumber \\
&&+2\kappa\sum_{m,n}\int dz\,\tilde \nu^2ag\epsilon_m \epsilon_n\int d^4x D_m^{\dagger}D_n
\eea
which suggests the normalization 

\be
\label{22X112}
2\kappa\int dz\, ag\epsilon_m \epsilon_n=\frac{m_n^2}{\tilde \nu^2}\delta_{mn}
\ee
as a result, (\ref{22X111}) takes  the canonical form for the free HL pseudoscalars
\be
\sum _n\int d^4x \left(\partial_{\mu}D_n^{\dagger}\partial^{\mu}D_n+m_n^2D_n^{\dagger}D_n\right)
\ee

\subsection{Heavy quark limit}

The heavy quark limit will be sought through the rescaling 
$z={\tilde z}/{\tilde \nu^\beta}$ with $\tilde \nu \rightarrow \infty$.
With this in mind  (\ref{22X7}) reads

\be
\label{22X11}
-\tilde\nu^{2\beta}\frac{g}{2f}\frac{d^2\phi_n}{d\tilde z^2}+\tilde\nu^{\beta}\frac{g^{\prime}}{2f}\frac{d\phi_n}{d\tilde z}+\tilde \nu^2a\phi_n=m_n^2\phi_n
\ee
with the limiting values

\bea
\label{22X12}
&&a\rightarrow a_0+\frac{a_0}{2}\left(\frac{\tilde z^2}{U_{KK}^2}\right)\frac{1}{\tilde \nu^{2\beta}}\nonumber\\
&&\frac{g}{2f},\frac{g^{\prime}}{2f}\rightarrow \frac {g_0}{2f_0}, \frac{g_0^\prime}{2f_0}+{\cal O}\left(\frac{1}{\tilde \nu^{\beta}}\right)
\eea
and $a_0=a(U_{KK})=(U_{KK}/R)^{\frac 32}$.  Notice that 

\be
\label{22X15}
\frac{d}{dz}\left(\frac{1}{2af}\frac{d(ag)}{dz}\right)\approx a_0+{\cal O}\left(\frac{1}{\tilde \nu^{2\beta}}\right)
\ee
This means that after the rescaling,  (\ref{22X10}) reduces to  (\ref{22X15}) as $\tilde\nu \rightarrow \infty$.
A consistent $\tilde \nu \rightarrow \infty$ limit is achieved by setting
$2\beta=2-2\beta\rightarrow \beta=\frac{1}{2}$ in (\ref{22X11}). Matching the leading poweres
in $\tilde\nu$ gives

\bea
\label{22XX14}
-\frac{g_0}{2f_0}\frac{d^2\phi_n}{d\tilde z^2}+\frac{a_0}{2}\frac{\tilde z^2\phi_n}{U_{KK}^2}=\tilde m_n^2\phi_n
\eea
with the squared mass

\bea
\label{22XX15}
m_n^2=\tilde \nu^2a_0+\tilde \nu \tilde m_n^2= M^2+\tilde \nu \tilde m_n^2
\eea
We have identified the heavy quark mass as  $M=\tilde\nu\sqrt{a_0}$.  
 More specifically, we can re-write the solutions to (\ref{22XX14}) as

\bea
\label{22XX16}
m_n^2
=&&M^2+\tilde\nu\sqrt{a_0}\left(\frac{g_0}{U_{KK}^2f_0}\right)^{\frac 12}\left(n+\frac 12\right)\nonumber\\
=&&M^2+M\left(\frac {2m_\rho^2}{0.67}\right)^{\frac 12}\left(n+\frac 12\right)
\eea
using the rho mass $m_\rho=\sqrt{0.67} M_{KK}$~\cite{SSX}. 
(\ref{22XX16}) is Regge-like with an intercept $M^2$ and a slope 
of about  $Mm_\rho$. Note that 
 the mass splitting $\Delta m$ between the odd-parity H-multiplet
 (say $n=0$) and the even-parity G-multiplet
(say $n=1$)  is finite in  this model, with 

\bea
\label{22X16}
\Delta m=m_{G}-m_{H}\approx \frac{m_\rho}{\sqrt{2}\sqrt {0.67}}
\eea
in the large $M$ limit. The splitting is about 
$\Delta m\approx  665$ MeV, which is larger than the reported mean 
of 420 MeV for charm and 396 MeV for bottom~\cite{CLEOII}. 

\section{HL correlation functions}

The HL correlation functions from the current-current correlator

\bea
\label{9X1}
{\Pi}^{AB}(q)=i\int d^4x e^{iq\cdot x} \left<0|T^*\left({\mathbb J}^A(x){\mathbb J}^B(0)\right)|0\right>
\eea
with the even and odd parity multiplet assigments

\bea
\label{9X}
({\mathbb J}^P, {\mathbb J}^V_\mu)=(\bar q i\gamma_5 Q, \bar q \gamma_\mu Q)\qquad
({\mathbb J}^S, {\mathbb J}^A_\mu)=(\bar qQ, \bar q i\gamma_\mu \gamma_5 Q)\nonumber\\
\eea
In the holographic approach,
(\ref{9X1}) can be obtained from the boundary effective action ${\bf S}_B$, by inserting sources in the UV and integrating out the
bulk fields using the equations of motion~\cite{HOLOXX,HOLOXXX, HOLOXXXX}.

\subsection{Vector and Axial-Vector  polarizations}

We now define  the bulk HL vector and scalar source fields in momentum space as

\bea
\label{XXX5}
&&\Phi_{\mu}(p,z)\rightarrow\frac{{\cal V}(p,z)}{{\cal V}(p,z_\Lambda)}\, {\mathbb S}_{\mu}(p)\nonumber\\
&&\Phi_z(p, z)\rightarrow 0
\eea
with $z_\Lambda/U_{KK}\gg 1$ setting the UV cutoff. The boundary-to-bulk propagator ${\cal V}(p,z)$ satisfies 
the off-shell version of (\ref{22X7}),

\be
\label{22X7X}
-\frac{d}{dz}  \left(  g  \frac{d {\cal V}}{dz}\right)+2f(p^2+\tilde \nu^2a){\cal V}=0
\ee
subject to  the axial and vector  boundary conditions

\bea
\label{XXX6}
{\cal V}(p,0)=&&0,\qquad\qquad axial\nonumber\\
\partial_z {\cal V}(p,0)=&&0,\qquad\qquad vector
\eea

To construct ${\bf S}_B$, we first make some general observations regarding the transverse eigenmode
equation (\ref{22X7X}-\ref{XXX6}).  In general, the equation admits two independent solutions. 
The first is  $f_1$ which is  square integrable
in the UV limit, and as $p^2\rightarrow -m_n^2$ , $f_1$ aproaches the normalized eigenmodes $\phi_n$ given in (\ref{22X7}). The
second is  $f_2$ which is another independent solution that  is not square integrable  in the UV,. We normalize it using the Wronskian
normalized Wronskian 

\be
\label{XXX1}
4\kappa g\left(f_2\frac{d}{dz}f_1-f_1\frac{d}{dz}f_2\right)=-1
\ee
Also, we note that near the UV boundary with $z\rightarrow \infty$, (\ref{22X7X}) simplifies to 

\be
\label{XXX2}
-\frac{d}{dz}\left(z^2\frac{d{\phi_n}}{dz}\right)+{\bf c}(p)\,z^{1\over 3}\phi_n\approx 0
\ee
with ${\bf c}(p)$ a p-dependent function. The two independent solutions to (\ref{XXX2}) take the asymptotic forms
in terms of the modified Bessel functions

\bea
\label{XXX3}
&&f_1(p,z)\approx \frac{K_3(6\sqrt{{\bf c}(p)}z^{\frac{1}{6}})}{\sqrt{z}}\rightarrow \frac{ e^{-6\sqrt{{\bf c}(p)}z^{\frac{1}{6}}}}{z^{\frac{7}{12}}}\equiv f_{1,asy}(p,z)
\nonumber\\
&&f_2(p,z)\approx \frac{I_3(6\sqrt{{\bf c}(p)}z^{\frac{1}{6}})}{\sqrt{z}}\rightarrow  \frac{ e^{6\sqrt{{\bf c}(p)}z^{\frac{1}{6}}}}{z^{\frac{7}{12}}}\equiv f_{2,asy}(p,z)\nonumber\\
\eea

The general  solutions to (\ref{22X7X}) 
satisfying  the boundary conditions (\ref{XXX6}) are 

\bea
\label{XXX7}
{\cal V}(p,z)=&&f_2(p,0)f_1(p,z)-f_1(p,0)f_2(p,z),\,\,axial\nonumber\\
{\cal V}(p,z)=&&f_2^{\prime}(p,0)f_1(p,z)-f_1^{\prime}(p,0)f_2(p,z),\,\,vector\nonumber\\
\eea
We now insert (\ref{XXX5}) using (\ref{XXX7}) into the massive quadratic action for $\Phi$. The result is the 
boundary action  ($z_\Lambda\rightarrow\infty$)

\bea
\label{XXX8}
{\bf S}_B=-\int \frac{d^4q}{(2\pi)^4}  {\mathbb S}^{\dagger}_{\mu}(q)
 \left(2\kappa   g(z_\Lambda)\frac{\partial_z{\cal V}(p,z_\Lambda)}{{\cal V}(p,z_\Lambda)}\right) {\mathbb S}^{\mu}(q)
\eea
from which we read the axial polarization function at the boundary

\bea
\label{XXX9}
\Pi_A(q^2)=
4\kappa g(z_\Lambda)\frac{f_2(q,0)f^{\prime}_1(q,z_\Lambda)-f_1(p,0)f_2^{\prime}(q,z_\Lambda)}{f_2(q,0)f_1(q,z_\Lambda)-f_1(q,0)f_2(q,z_\Lambda)}\nonumber\\
\eea
 In the vicinity 
of the poles,  (\ref{XXX9}) is dominated by

\be
\label{XXX10}
\Pi_A(q^2)\rightarrow -\frac{f_2(q,0)}{f_1(q,0)}4\kappa g(z_\Lambda)\frac{f_1^{\prime}(q,z_\Lambda)}{f_2(q,z_\Lambda)}
\ee
To simplify (\ref{XXX10}), we now note that  near the poles the identity (\ref{XXX1}) simplifies

\be
\label{XXX11}
4\kappa g\frac{f_1^{\prime}}{f_2}=-\frac 1{f_2^{2}}\rightarrow -\frac 1{f_{2,asy}^{2}}
\ee
where the last relation holds at  the UV boundary $z=z_\Lambda$. 
(\ref{XXX11}) when used in (\ref{XXX10}) reduces the axial polarization function to 

\be
\label{XXX12}
\Pi_{A}(p^2)=\frac{f_2(p,0)}{f_1(p,0)}{f^{-2}_{2,asy}(p,z_\Lambda)}
\ee
A similar reasoning with the vector sources gives the vector polarization function

\be
\label{XXX13}
\Pi_{V}(q^2)=\frac{f_2^{\prime}(q,0)}{f_1^{\prime}(q,0)}{f^{-2}_{2,asy}(q,z_\Lambda)}
\ee
The poles in the axial-vector correlator (\ref{XXX12}) are given by $f_1(m_n,0)=0$,
while those of the vector correlator (\ref{XXX13}) are given by $f^\prime_1(m_n,0)=0$,
in agreement with the mass spectrum in (\ref{22X7}). The residues are sensitive to
the UV cutoff $z_\Lambda$. They are further discussed in the Appendix.

\subsection{Scalar and Pseudo-Scalar polarizations}

Similarly, for the scalar and pseudo-scalar mesons, we refer to the square integrable solutions by $\tilde f_1$
and to the non-square integrable function by $\tilde f_2$, and require the normalization through the Wronskian

\be
\frac{\kappa\tilde \nu^2}{p^2}\frac {\tilde g}{\tilde f} \left(\tilde f_1\frac{d\tilde f_2}{dz}-\tilde f_2\frac{d\tilde f_1}{dz}\right)=-1
\ee
in comparison to the vector normalization in (\ref{XXX1}). A repeat of the arguments for the vector
polarizations lead to the pseudo-scalar and scalar polarizations 

\bea
&&\tilde \Pi_{S}(q^2)=\frac{\tilde f_2(q,0)}{\tilde f_1(q,0)}{\tilde f^{-2}_{2,ayy}(p,z_\Lambda)}\nonumber\\
&&\tilde \Pi_{P}(q^2)=\frac{\tilde f_2^{\prime}(q,0)}{ \tilde f_1^{\prime}(q,0)}{\tilde f^{-2}_{2,ayy}(q,z_\Lambda)}
\eea
respectively. It is readily checked that in the heavy quark limit $\Pi=\tilde\Pi$, and the 
${\mathbb J}\mathbb J$ correlators exhibit heavy-quark symmetry. This degeneracy follows from the
rigid $O(4)$ symmetry of the vector fields in 5-dimensions.


\section{HL chiral interactions}

\subsection{Chiral symmetry }

To identify properly the nature of the chiral transformation on the holographic field decomposition,
we will recall in this section how this identification is made in the constituent quark model. For that
consider the HL sources $H_\pm$  in the effective action for bare constituent quarks 

\be
-\bar \psi_{L} H_{+}Q_{R}-\bar \psi_{R}H_{-}Q_{L}+{\rm c.c}
\ee  
with $H_\pm$ HL sources for the $(0^\pm,1^\pm)$ multiplet with $\pm$ chiralities. 
Under rigid chiral symmetry,  the pion field $U\rightarrow LUR^\dagger$, and the fundamental quarks 
transform as $\psi_{L}\rightarrow L\psi_{L},\psi_{R}\rightarrow R\psi_{R}$,  so that

\be
(H_+, H_-)\rightarrow (LH_+, RH_-)
\ee
Rigid chiral symmetry is better enforced through the decomposition $U=\xi_L\xi_R^\dagger$
with $\xi_{L}\rightarrow L\xi_{L}h^{\dagger}(x)$ and $\xi_{R}\rightarrow R\xi_Rh^\dagger (x)$.
Consider now the dressed constituent quarks $\chi_{L,R}=\xi_{L,R}^{\dagger}\psi_{L,R}$.
It follows, that the corresponding dressed HL effective fields with odd parity are

\be
\label{HDD}
H=({\gamma_{\mu}D^{\mu}+i\gamma_{5}D})
\ee
with the identification

\bea
D^{\mu}=&&\xi_{L}^{\dagger}H^{\mu}_{+}+\xi_{R}^{\dagger}H_{-}^{\mu}\nonumber\\
D=&&\xi_{L}^{\dagger}H_{+}+\xi_{R}^{\dagger}H_{-}^{\mu}
\eea
Under chiral symmetry (\ref{HDD}) transforms as $H\rightarrow h(x)H$. Similarly,
for the $(0^+,1^+)$ multiplet we can define the dressed HL effective fields with even parity as

\be
G=(D_0+\gamma_\mu\gamma_5D^\mu_1)
\ee
which transforms as $G\rightarrow h(x)G$ under chiral symmetry
We now seek to enforce these symmetries on the bulk fields in holography.

\subsection{Hologaphic identification}

In the axial gauge  $A_{M}(x,z\rightarrow \infty)\rightarrow 0$, the residual gauge transformation $g$ satisfies
 $\partial_{M}g\rightarrow 0$ at infinity. Following the arguments in \cite{SSX},  we identify the rigid $L,R$
 chiral transformations as $L=g(z\rightarrow +\infty)=g_+$ and $R=g(z\rightarrow -\infty)=g_-$. In the axial gauge (bare gauge),
 the pion field $\pi(x)$ is identified as
 
 \be
 \label{AZZ}
 U(x)=e^{\frac{i}{f_\pi}\pi (x)}=Pe^{-\int_{-\infty}^{+\infty}A_z(x,z^\prime)dz^{\prime}}
 \ee
As expected, under rigid chiral transformations $U\rightarrow LUR^{\dagger}$. A useful gauge choice
is the one where the pion field is identified with the zero mode in the holographic direction (dressed gauge)

\be
\label{AZZX}
A_{z}(x,z)=-\frac{i}{f_{\pi}}\pi(x)\psi_{0}^{\prime}(z),\qquad A_{\mu}(x,z)=0
\ee
The non-normalizable pion zero-mode running along the $(N_f-1)$ coincidental $D8_L$ branes is~\cite{SSX}

\be
\psi_0(z)=\frac 1\pi {\rm arctan}\left(\frac z{U_{KK}}\right)
\ee
This dressed gauge is reached by noting that 
the pion identification in (\ref{AZZ}) allows us to slice the holonomy along the z-direction
through 

\be
\xi_{\pm}(x,z)=Pe^{-\int_{0}^{\pm z}A_{z}(x,z^{\prime})dz^{\prime}}
\ee
and identify

\bea
\xi_{L}(x)=&&\xi_{+}(x,z\rightarrow\infty)\nonumber\\
\xi_{R}^{-1}(x)=&&\xi_{-}(x,z\rightarrow \infty)
\eea

The bare HL meson fields transform as fundamental fields under rigid chiral
transformations

\be
\Phi(x,z\rightarrow \pm \infty)\rightarrow g_{\pm} \Phi (x,z\rightarrow \pm \infty)
\ee
This means that the bare $\Phi(z\rightarrow \pm \infty)$  are the analogue of the bare $H_{\pm}$. 
In particular for charm, the low lying $(\tilde\phi_0, \phi_0)$ modes contribute to the $(D,D^*)$ meson multiplet $H$, and the first excited 
$(\tilde\phi_1,\phi_1)$ modes contribute to the $(D_0,D_1)$ meson  multiplet G. The dressed HL meson fields are readily identifiedas
\bea
\label{BX1}
\Phi_{\mu}(x,z)=&&\xi_{+}(x,z)\phi_0(z)D_{\mu}(x)\nonumber\\
\Phi_{z}(x,z)=&&\xi_{+}(x,z)\tilde \phi_0(z)D(x)
\eea

\subsection{Quadratic HL holographic action}

The holographic effective action with all quadratic terms in the HL fields can now be constructed
without recourse to the heavy quark limit. For that, we follow the construction in 
\cite{SSX} and supplement the $A_M$ field with external 
flavor sources $A_{L,R}$ by defining

\bea
\label{XX1}
A_{\mu}=V_{\mu}+2\psi_0A_{\mu}+\sum_{n}{\mathbb A}_{\mu,n}\psi_{n}
\eea
with $A_z$ still given by  (\ref{AZZX}).
Here, the external sources are defined as 
$A_{L,R}=(V\pm A)$. For odd values of $n$ we identify ${\mathbb A}_{\mu,n}=v_{\mu, n}$
with the light $1^{--}$ flavor vector excitations,  and for even values of $n$ we identify ${\mathbb A}_{\mu,n}=a_{\mu, n}$
with the light $1^{++}$ flavor axial excitations. Using the additional definition

\be
\label{XX2}
{\bf A}_{\mu}(x,z)=e^{-\frac{i}{f_{\pi}}\psi_0\pi}(\partial_{\mu}+A_{\mu})e^{\frac{i}{f_{\pi}}\psi_0\pi}
\ee
we have the identities

\bea
\label{XX3}
&&e^{-\frac{i}{f_{\pi}}\psi_0\pi}f_{\mu,\nu}=\sum_{n}\phi_n(\partial_{[\mu}D_{n,\nu]}+{\bf A}_{[\mu}D_{n,\nu]})\nonumber\\
&&e^{-\frac{i}{f_{\pi}\psi_0}\pi}f_{\mu,z}=\sum_{n}\left(\tilde \phi_n(\partial_{\mu}D_n+{\bf A}_\mu D_n)
-\phi^{\prime}_nD_{\mu,n}\right)\nonumber\\
\eea

The DBI contributions which are quadratic in $D,D^*$ and linear in ${\bf A}$ follows by inserting (\ref{XX1}-\ref{XX3}).
The resulting action is of the form $S_{2}=\kappa f{\bf  S}_1+\kappa g{\bf S}_2$ with 

\bea
\label{XX4}
&&{\bf S}_1=\nonumber \\ &&+4\sum_{m,n}\phi_m\phi_nD_m^{\mu^{\dagger}}(\partial_{\mu}{\bf A}_{\nu}-\partial_{\nu}{\bf A}_{\mu}+[{\bf A}_{\mu},{\bf  A}_{\nu}])D_n^{\nu}\nonumber \\ &&-4\sum_{m,n}\phi_m\phi_nD_{m}^{\dagger \mu}({\bf A}_{\nu}{\bf A}_{\mu}-h_{\mu \nu}{\bf A^2})D_{n}^{\nu}\nonumber \\&&-2\sum_{mn}\phi_m\phi_n\partial_{[\mu}D_{n,\nu]}^{\dagger}{\bf A}^{[\mu}D_m^{\nu]}\nonumber \\&&-2\sum_{m,n}\phi_m\phi_nD_m^{[\mu\dagger}{\bf A}^{\nu]}\partial_{[\mu}D_{n,\nu]}\\
&&{\bf  S}_2=\nonumber \\ &&-2\sum_{m,n}\tilde \phi_m\tilde \phi_n(\partial_{\mu}D_m^{\dagger}{\bf A}^{\mu}D_{n}-D_n^{\dagger}{\bf  A}^{\mu}\partial_{\mu}D_m)\nonumber \\ &&+2\sum_{m,n}\tilde \phi_m\phi^{\prime}_n(D^{\mu\dagger}_n{\bf A}_{\mu}D_m-D_m^{\dagger}{\bf A}_{\mu}D_n^{\mu})\nonumber \\&& +2\sum_{m,n}\tilde \phi_m\phi_n(D_m^{\dagger}\partial_z{\bf  A}_\mu D_n^{\mu}-D_n^{\mu\dagger}\partial_z{\bf A}_{\mu}D_m)\nonumber \\ &&+2\sum_{m,n}\tilde \phi_m\tilde \phi_nD_m^{\dagger}{\bf A^2}D_n
\eea
where we have omitted the traces and the 
integrations for notational simplicity. Here we have defined ${\bf A^2}={\bf A}^{\mu}{\bf  A}_{\mu}$.
The expansion of ${\bf A}_\mu$ in terms of the pion and vector mesons in leading orders, read

\bea
\label{XX5}
&&{\bf A}_{\mu}=\frac{i}{f_{\pi}}\psi_0\partial_{\mu}\pi+\frac{1}{8f_{\pi}^2}[\pi,\partial_{\mu}\pi]+...\nonumber \\
&&+V_{\mu}-\frac{i}{f_{\pi}}\psi_0[\pi,V_{\mu}]+2A_{\mu}\psi_0-\frac{2i}{f_{\pi}}\psi_0^2[\pi,A_{\mu}]+...\nonumber \\
&&+\sum_{n}\psi_{2n-1} v_{n,\mu}+\sum_n\psi_{2n}a_{n,\mu}+...
\eea

The unexpanded CS contribution involving the $\Phi$ fields receives several contributions from (\ref{6}).
They are

\be
\label{CSXX1}
S_{CS}=S_{\Phi^2,A}+S_{\Phi^2,A^2}+S_{\Phi^2,A^3}+ S_{\Phi^4,A}+S_{\Phi^4}
\ee
with each of the contributions given in form-notations as follows

\bea
\label{CSXX2}
S_{\Phi^2,A}=&&-\frac{N_c}{24\pi^2}(d\Phi^{\dagger}Ad\Phi+ d{\Phi}^{\dagger}dA\Phi+\Phi^{\dagger}dAd\Phi)\nonumber\\
S_{\Phi^2,A^2}=&&-\frac{N_c}{16\pi^2}(d{\Phi}^{\dagger}A^2\Phi+\Phi^{\dagger}A^2d\Phi)\nonumber \nonumber\\
&&-\frac{N_c}{16 \pi^2}\Phi^{\dagger}(AdA+dAA)\Phi\nonumber\\
S_{\Phi^2,A^3}=&&-\frac{5N_c}{48\pi^2}\Phi^{\dagger}A^3\Phi\nonumber\\
S_{\Phi^4,A}=&&+\frac{N_c}{8\pi^2}\Phi^{\dagger}\Phi\Phi^{\dagger}A\Phi\nonumber\\
S_{\Phi^4}=&&+\frac{N_c}{16\pi^2}\Phi^{\dagger}\Phi(\Phi^{\dagger}d\Phi+d\Phi^{\dagger}\Phi)
\eea
The $\Phi$ field is defined explicitly in (\ref{BX1})  and the $A$ field is defined in (\ref{XX1}). We have omitted
the flavor trace and the integration which is 5-dimensional here. The latter will reduce to 4-dimensions after inserting (\ref{BX1})
and integrating over the HL meson holographic wavefunctions.

We note that when only the pion field is retained in 
(\ref{XX1}) (no vector mesons), (\ref{CSXX1}-\ref{CSXX2}) simplifies as 

\be
S_{\Phi^2,A^2}+S_{\Phi^2,A^3}\rightarrow 0
\ee
Also, the first quadratic contribution $S_{\Phi^2,A}$ does not vanish and will be discussed in details below.
In addition, the quartic contributions in (\ref{CSXX2})  do not vanish and contribute

\bea
&&S_{\Phi^4,A}+S_{\Phi^4} \rightarrow\nonumber\\
&&\frac{iN_c}{8\pi^2}D^{\dagger}DD^{\dagger}{\bf A}D+\frac{iN_c}{16\pi^2}D^{\dagger}D(D^{\dagger}dD+dD^{\dagger}D)
\eea

Finally, the DBI quadratic holographic action (\ref{XX4}) together with the CS parts (\ref{CSXX2}) 
are  the most general pion and vector meson interactions
with HL mesons with finite masses. The HL mesons are characterized by a pionic-like zero mode as noted in (\ref{22X7}) 
in the massless limit. So (\ref{XX4}) and (\ref{CSXX2}) 
interpolate continuously between massless and massive HL mesons with exact heavy
quark symmetry asymptotically as we further detail  below.

\subsection{One-pion interaction}

Now we consider the  sourceless 
case with $A_{L,R}=0$ and all $a_{n,\mu},v_{n,\mu}=0$. In this case, in leading order in the pion field, the bare heavy meson fields in 
(\ref{BX1}) reads

\bea
&&\Phi_{\mu}\approx \left(1+\frac{i}{f_{\pi}}\psi_{0}\pi\right)\phi_1D_{\mu}\nonumber\\
&&\Phi_{z}\approx \left(1+\frac{i}{f_{\pi}}\psi_{0}\pi\right)\tilde \phi_1D\nonumber\\
&&A_{z}=-\frac{i}{f_{\pi}}\pi\psi_0^{\prime}(z),\,\,\,A_{\mu}=0
\eea
in terms of which the contributions (\ref{2XX2}-\ref{2XX3}) are

\bea
&&f_{\mu\nu}\approx \nonumber \\ &&\left(1+\frac{i}{f_{\pi}}\psi_{0}\pi\right)\phi_1\partial_{[\mu}D_{\nu]}+\frac{i}{f_{\pi}}\psi_{0}\phi_1\partial_{[\mu}\pi D_{\nu]}\nonumber\\
&&f_{\mu z}^{0}\approx \nonumber \\ &&\left(1+\frac{i}{f_{\pi}}\psi_{0}\pi\right)\tilde  \phi_1\partial_{\mu}D-\frac{i}{f_{\pi}}\pi \psi_0 \tilde \phi_1^{\prime }D_{\mu}\nonumber \\&&+\frac{i}{f_{\pi}}\psi_{0}\tilde \phi_1\partial_{\mu}\pi D
\eea
and

\bea
&&a_{\mu\nu}\approx \left(1+\frac{i}{f_{\pi}}\psi_{0}\pi\right)D_{[\mu}D^{\dagger}_{\nu]}\left(1-\frac{i}{f_{\pi}}\psi_{0}\pi\right)\phi_1^2\nonumber\\
&&a_{\mu z}\approx \left(1+\frac{i}{f_{\pi}}\psi_{0}\pi\right)D_{[\mu}D^{\dagger}_{z]}\left(1-\frac{i}{f_{\pi}}\psi_{0}\pi\right)\phi_1\tilde \phi_1\nonumber\\
\eea
with 

\bea
F_{\mu \nu}={\cal O}(\pi^2), \qquad
F_{z \mu}\approx \psi^{\prime }_{0}\frac{i}{f_{\pi}}\partial_{\mu} \pi
\eea

In developing the gauge  and heavy meson fields in 
(\ref{BX1}), we have omitted the contributions from the tower of vector  and axial fields,
and the contribution of the excited heavy mesons for simplicity. They will be recalled below. 
Note that in the dressed gauge, the leading pion contribution
is the  current algebra result

\bea
\label{3X2}
\frac {f_\pi^2}{4}{\rm Tr}(U^{-1}\partial_{\mu}U)^2
\eea
with $f_\pi^2/M_{KK}^2=4\kappa/\pi$.

\subsubsection{$(0^-,1^-)$ multiplet}

The leading contribution to the interaction between the heavy-light mesons to the pions follows only from the 
first contribution in the CS term in (\ref{6}) in the form

\be
\label{4X1}
S_{\rm CS}\rightarrow -\frac{N_c}{16\pi^2}\int dz d^4x {\rm Tr}(d\Phi^{\dagger}dA\Phi+\Phi^{\dagger}dAd\Phi)
\ee
Using the identification (\ref{HDD}), we can re-write (\ref{4X1}) as 


\bea
\label{4X4}
S^\pi_{\rm CS}= &&-\frac{N_c}{32\pi^2f_{\pi}}\int dz \psi_0^{\prime}(z)\phi_0^2 \nonumber \\ 
&&\times\int d^4x{\rm {\bf Tr}}\left( \gamma \cdot \partial\bar H\gamma_5\gamma \cdot \partial\pi H +{\rm c.c.}\right)
\eea
with $\bar H=\gamma^0H^\dagger \gamma^0$. We note that the bolded trace in (\ref{4X4}) is now both over flavor and spin.
To obtain the non-relativistic reduction of (\ref{4X4}), we first 
decompose the positive frequency part of $H\rightarrow {\mathbb H}_++{\mathbb H}_-$ as

\be
\label{HNR}
{\mathbb H}_{\pm}=\frac{e^{-iMx_0}}{\sqrt{2M}}(\gamma_{\mu}D_{\mu}+i\gamma_5D) \frac{1\pm\gamma_0}{2}
\ee
which gives 

\bea
\label{4X44}
&&S^\pi_{\rm CS}=S_{\rm CS}^++S_{\rm CS}^- \nonumber \\ 
&&S_{\rm CS}^{\pm}=
\frac{iN_c}{32\pi^2f_{\pi}}\int dz \psi_0^{\prime}(z)\phi_0^2\nonumber \\
&&\times \int d^4x \,
{\rm Tr}\partial_{i}\pi (\pm (D_{i}D^{\dagger}-DD^{\dagger}_i)+\epsilon^{ijk}D_{k}D_{j}^{\dagger})
\eea
Keeping only the contribution from ${\mathbb  H}_{+}$, which transforms homogeneously
as the $(\frac 12, \frac 12)$ representation under heavy-light spin transformations, is equivalent to deforming the 
CS  contribution (\ref{4X4}) to

\be
\label{DEF}
&&S^\pi_{\rm CS}\rightarrow S^+_{\rm CS} =-\frac{N_c}{32\pi^2f_{\pi}}\int dz \psi_0^{\prime}(z)\phi_0^2\nonumber\\ &&\times\int d^4x
{\rm {\bf Tr}}\left( (\gamma \cdot \partial+iM)\bar H\gamma_5\gamma \cdot \partial\pi H +{\rm c.c.}\right)\nonumber\\
\ee 
Amusingly, this  deformation can be viewed as a fermion loop with a  massive instead of a massless quark.
Therefore, keeping positive energy naturally selects the ${\mathbb H}_{+}$ contribution. The modified term is actually a normal term which requires a metric and may come from a missing piece of the low-energy effective field theory action of our underlying brane configuration with large transverse separation. Assuming this, the z-integration can be performed exactly to give

\bea
\label{4XX4}
&&\int dz \psi_0^{\prime}(z)\phi_0^2=\frac{1}{\pi U_{KK}}\frac{1}{4 f_0\kappa} =\nonumber\\
&&\frac{1}{ \pi \tilde T(2\pi\alpha^{\prime})^2R^3}=\frac{108\pi^3l_s^2}{M_{KK}N_c}\times  \frac{2M_{KK}}{\pi \lambda l_s^2}=\frac{216\pi^2}{N_c\lambda}\nonumber\\
\eea
which allows for (\ref{DEF}) to take the standard leading one-pion interaction form

\bea
\label{4X444}
S^+_{\rm CS}=\frac{g_H}{f_\pi}
 \int d^4x \,{\rm Tr}\partial_{i}\pi
(D_{i}D^{\dagger}-DD^{\dagger}_i+\epsilon^{ijk}D_{k}D_{j}^{\dagger})
\eea
with the pseudo-vector pion  axial coupling $g_H$ to the $(0^-,1^-)$ multiplet given by

\be
g_H\equiv \frac{27}{4\lambda}
\ee
In~\cite{SSX} a fit to the low lying meson spectrum led to $\lambda\sim 8.7$ which implies that $g_H\sim 0.78$
in our holographic model. The holographic result is close to the reported value of $g_H\sim 0.65$,
as measured through the charged pion decay $D^*\rightarrow D\pi$~\cite{CLEOII}.

\subsubsection{$(0^+,1^+)$ multiplet}

Using the analogue of the non-relativistic
reduction (\ref{HNR})  for the G-multiplet with

\be
\label{GNR}
G\rightarrow {\mathbb G}_+=\frac{e^{-iMx_0}}{\sqrt{2M}}(D_0+\gamma_{\mu}\gamma_5D_1^{\mu}) \frac{1+\gamma_0}{2}
\ee
a similar reasoning shows that  the pseudo-vector pion  axial coupling $g_G$ 
to the $(0^+,1^+)$ multiplet follows similarly with the mode substitution $\phi_0\rightarrow \phi_1$
in (\ref{4X4}-\ref{4XX4}). As a result, we have $g_G=g_H$ in our holographic construct.

The intra-multiplet one-pion interaction is now seen to follow from the DBI contribution only,

\bea
\label{HGHG}
&&S^\pi_{HG}= \nonumber\\
&&+\frac{4\kappa}{f_{\pi}}\int f\phi_0\phi_1\psi_{0}(z)dz \int d^4x {\rm Tr}\partial_{0}\pi(D_{i}D^{\dagger}_{1i}+D_{1i}D^{\dagger}_{i})\nonumber \\&&+\frac{2\kappa}{f_{\pi}}\int g\tilde  \phi_0\tilde \phi_1\psi_{0}(z)dz 
\int d^4x {\rm Tr}\partial_{0}\pi(DD_0^{\dagger}+D_0D^{\dagger})\nonumber\\
\eea
We note the following identity 

\be
2\int dz\,f \psi_{0}\phi_0\phi_1=\int dz\, g\psi_{0}\tilde \phi_0\tilde \phi_1
\ee
which allows to rewrite (\ref{HGHG}) in standard form 

\be
\label{GHH}
\frac{g_{HG}}{f\pi}{\rm {\bf Tr}}\left(\gamma_5\bar G Hv^{\mu}A_{\mu}\right)+{\rm c.c}
\ee
The pseudo-vector axial cross pion coupling is

\be
\label{GHGHGH}
g_{HG}=4\kappa \int dz\, f\psi_{0}\phi_0\phi_1=\frac{2^{\frac 14}}{2\pi}\left(\frac{M_{KK}}{M}\right)^{\frac 12}
=0.23\left(\frac{m_\rho}{M}\right)^{\frac 12}\nonumber\\
\ee
the last relation follows from the substitution of the rho mass $m_\rho=\sqrt{0.67} M_{KK} $~\cite{SSX}.
The axial coupling is seen to vanish in the heavy quark limit $M=\tilde\nu\sqrt{a_0}\rightarrow \infty$.

We observe that in the present holographic model all one-pion couplings to the  HL mesons are pseudo-vectors.
In particular, the Goldberger-Treiman combination $g_{GH}\Delta m/f_\pi$ does not  support a pseudo-scalar coupling 
as initially noted in the chiral symmetric constructs without confinement in~\cite{MACIEK,BARDEEN}.
Using the empirical value of $m_\rho\sim 770$ MeV and the charm mass with $m_C=1275$ MeV we
find that $g_{HG,C}\sim 0.18$,  while for bottom with $m_B\sim 4180$ MeV we find $g_{HG,B}\sim 0.10$.

\subsubsection{One-pion radiative widths}

The strong intra-multiplets decay $G\rightarrow H+\pi$ follow from (\ref{GHH}). Both the chargeless and charged 
pion decay of charmed and bottom mesons with final momentum $p_\pi$ read

\bea
&&\Gamma(G\rightarrow H+\pi^0) =\frac 1{4\pi}\left(\frac{g_{HG,C}}{f_\pi}\right)^2 (m_G-m_H)^2|p_{\pi^0}|\nonumber\\
&&\Gamma(G\rightarrow H+\pi^\pm) =\frac 2{4\pi}\left(\frac{g_{HG,C}}{f_\pi}\right)^2 (m_G-m_H)^2|p_{\pi^\pm}|\nonumber\\
\eea
Our holographic result for the HL charm meson gives

\bea
&&\Gamma (D_1^0(2420)\rightarrow D^{*+}(2010) \pi^-)=\nonumber\\
&&\frac 1{2\pi}\left(\frac{0.18}{93}\right)^2(411)^2(354)=36\, {\rm MeV}
\eea
wich is comparable to the measured full width of $=(27.4\pm 2.5)$ MeV  at $p_{\pi^-}=354$ MeV~\cite{PDG}. The 
partial width for the charged decay of the HL bottom meson is

\bea
&&\Gamma (B_1^0(5721)\rightarrow B^{*+}(5325) \pi^-)=\nonumber\\
&&\frac 1{2\pi}\left(\frac{0.1}{93}\right)^2(400)^2(362)=11\, {\rm MeV}
\eea
which is to be compared to a full width of $23\pm 5$ MeV at $p_{\pi^-}=362$ MeV~\cite{PDG}. Also, we 
have the partial decay width

\bea
&&\Gamma (B_1^+(5721)\rightarrow B^{*0}(5325) \pi^+)=\nonumber\\
&&\frac 1{2\pi}\left(\frac{0.1}{93}\right)^2(402)^2(409)=12\, {\rm MeV}
\eea
which is to be compared to the measured full width of $49^{+12}_{-16}$ MeV at $p_{\pi^+}=409$ MeV~\cite{PDG}.

\subsection {Two pion interaction}

The two pion interaction terms can be derived using similar arguments.  All terms quadratic in $D,D^*$ and linear in ${\bf A}$
following from the DBI action (\ref{XX4})  follow from the two contributions

\bea
\label{T1}
&&-4i\sum_{m,n}f\phi_m\phi_n\,{\rm Tr}({\bf A}_0D^{\mu}_mD^{\dagger}_{\mu n})\nonumber\\ 
&&-2i\sum_{m,n} g \tilde \phi_m\tilde \phi_n\, {\rm Tr}({\bf A}_0D_mD_n^{\dagger})
\eea
Inserting (\ref{XX5}) in (\ref{T1}) and performing the z-integrations give

\be
\label{T2}
-\frac{i}{8f_{\pi}^2}\sum _{n}{\rm Tr}\left([\pi,\partial_{0}\pi](D_{n,\mu}D_{n}^{\mu\dagger}+D_nD_n^{\dagger})\right)
\ee
which is of order $M^0$. For instance, 
(\ref{T2}) describes the parity conserving two pion radiative decays  within the same multiplets $(0^\pm, 1^\pm)$,
i.e. $H\rightarrow H+2\pi$ and $G\rightarrow G+2\pi$.

In addition, there are also parity non-conserving 
two pion interactions from the Chern-Simons term, 
leading to radiative decays of the type  $G\rightarrow H+2\pi$. To leading order in the heavy-quark limit,
we have

\bea
\label{T3}
&&\frac{1}{8f_{\pi}^2M}\sum_{m,n}\int dz\, \kappa g\tilde \phi_m\phi^{\prime}_n\int d^4x \nonumber \\ 
&&\times {\rm Tr}\left([\pi,\partial_i \pi](D_mD_{ni}^{\dagger}-D_{n,i}D_m^{\dagger}+\epsilon^{ijk}D_{mk}D_{nj}^{\dagger})\right)\nonumber\\
\eea
with 

\bea
\label{T4}
\int dz \,\kappa g\tilde \phi_0\phi^{\prime}_1=\left(\frac{g_0}{f_0}\right)^{1/2}\int dz\, \kappa f\phi_0\phi_1^{\prime}=\nonumber \\
 \frac{\sqrt{2}}{4}\left(\frac{g_0}{f_0}\right)^{1\over 4}\left(\frac{M}{U_{KK}}\right)^{1\over 2}=
\frac{2^{3\over 4}}4\left(M_{KK}M\right)^{1\over 2}
\eea
(\ref{T4}) yields the two-pion cross coupling $G_{GH}/f_\pi^2$ in (\ref{T3}) as 

\be
\label{T5}
G_{GH}=\frac{2^{3\over 4}}{32}\left(\frac{M_{KK}}{M}\right)^{1\over 2} = 0.06 \left(\frac{m_\rho}{M}\right)^{1\over 2}
\ee
which is relatively small.
The couplings $G_{GH}$ in (\ref{T5}) and $g_{GH}$ in (\ref{GHGHGH}) are both of order $1/\sqrt{M}$ and vanish in the 
heavy quark limit.

\section{Conclusions}

We have presented a top-down holographic approach to the HL mesons interacting with the lightest 
pseudoscalar mesons. The geometrical set up consists of $N_f-1$ light $D8$-$D\bar 8$  probe branes plus
one heavy brane   in the background  of $N_c$ $D4$ branes. We have identified the HL  degrees 
of freedom with the string low energy degrees of freedom near the world volume of the light branes. They are
represented by bi-fundamental vector fields that are approximately local in the vicinity of the light branes.
Unlike any other model of QCD, the holographic set up confines 
in the infrared and exhibits the spontaneous breaking of chiral symmetry.

We have shown how the holographic effective action emerges from the bulk DBI and Chern-simons actions,
and explicitly verified that it enjoys both chiral and heavy quark symmetry in the limit of a heavy quark mass,
modulo  a suitable deformation of the CS contribution. 
The HL holographic effective action reduces continuously to the chiral effective action
for $SU(N_f)$ when the heavy quark mass is removed in the limit of coincidental $N_f$ light branes. 
 This construction can be made  more realistic
 through the use of improved holographic QCD~\cite{KIRITSIS}.

The HL effective action allows for a description of the HL meson internal structure in the strong
coupling $\lambda=g^2N_c$ limit. In particular, the squared mass spectrum is shown to be Regge-like 
with fixed intercept $M^2$ and a slope of about $Mm_\rho$. In leading order, the splitting between 
the even and odd parity multiplets
is fixed by the rho mass $m_\rho$. 

We have made explicit use of the HL effective action to extract the pertinent axial charges for the 
low lying HL multiplets $H=(0^-,1^-)$ and $G=(0^+,1^+)$ both of which are degenerate
in the heavy quark limit. Holography shows that the axial couplings are equal with $g_H=g_G=27/4\lambda$, and
close to the experimentally reported value of $g_H=0.65$,  for $\lambda\sim 8.7$ which is
the value selected in~\cite{SSX}. The inter-multiplet
coupling is fixed by the ratio of the rho to heavy quark mass as  $g_{GH}=0.23\,(m_\rho/M)^{1\over 2}$.
The ensuing one-pion partial decay widths are consistent with the total decay widths
 reported by the recent measurements  for both charm and bottom without any adjustment.  
 
 The shortcomings of the holographic limit are rooted in the double limit of large $N_c$ and strong 
 $^\prime$t Hooft coupling $\lambda=g^2N_c$. The corrections are notoriously hard to calculate.
 This notwithstanding, most of the leading order results we have established are close to the empirical
values, including the recently reported charged pion decays for HL bottom mesons. The exceptions are
the HL decay constants based on our  stringy estimate of the heavy mass as detailed in the Appendix.  
The holographic HL chiral effective action provides  valuable results
 for the few pion decays without and with vector mesons that could be compared with the upcoming
 experiments involving especially HL bottom mesons. It also provides  a framework for discussing
 electromagnetic decays, as well as single and double heavy baryons. Some of these issues will
 be addressed next.

\section{Acknowledgements}

We thank Rene Meyer for a discussion. 
This work was supported by the U.S. Department of Energy under Contract No.
DE-FG-88ER40388.

\section{Appendix: HL decay constants}

The axial and vector polarization functions (\ref{XXX12}-\ref{XXX13}) exhibit poles, with the squared axial decay 
constants  as residues

\be
f_{A_n,V_n}^2=-\lim_{q^2\to -m_n^2}\left(\left({q^2}+{m_n^2}\right)\,\Pi_{A,V}(q^2)\right)
\ee
In the heavy mass limit, the residues at the poles of the scalar and pseudoscalar
polarization functions (\ref{XXX12}-\ref{XXX13})  are related 
by heavy quark symmetry which holds in our case. 
In the standard definitions, the pseudoscalar constant is $m_Pf_P\equiv f_A$
and vector decay constant is $m_{P^*}f_{P^*}\equiv f_V$, with 

\bea
\label{DEXX}
\left<0|\bar q i\gamma_\mu \gamma_5 Q|P(p)\right>=&&f_Pp_\mu\nonumber\\
\left<0|\bar q \gamma_\mu  Q|P^*(p)\right>=&&f_{P^*}\epsilon_\mu
\eea
for $P=D,\bar B, ...$ and $P^*=D^*,\bar B^*,...$ respectively. (\ref{DEXX}) are 
measurable through the weak leptonic decays $D\rightarrow \bar l\nu_l$ and $\bar B\rightarrow l\bar\nu_l$.

\subsubsection{Generalities}

To explicit the decay constants as residues at the poles of polarization functions derived above,
we first reorganize the equation for the bulk-to-boundary propagator ${\cal V}(p,z)$ in the heavy
quark limit using (\ref{22XX14}), 

\be
\label{ZZ1}
-\frac{d^2{\cal V}}{\omega_0d\tilde z^2}+(\sqrt{\omega_0}\tilde  z)^2{\cal V}=-\frac{2(p^2+M^2)f_0}{\omega_0\tilde \nu g_0}{\cal V}
\ee
with $\omega_0^2={f_0a_0}/{g_0U_{KK}^2}$. We can set 

\bea
\label{ZZ2}
\alpha\equiv -\frac{p^2+M^2}{\omega_0\tilde \nu}\frac{f_0}{g_0}-\frac{1}{2}\rightarrow
\frac{\tilde m_n^2}{\omega_0}\frac{f_0}{g_0}-\frac{1}{2}
\eea
where the last identity holds for $p^2=-m_n^2$. Using 
 the parametrization $x=\sqrt{2\omega_0}\tilde z=\sqrt{2\omega_0\tilde \nu}z$, we can re-write
(\ref{ZZ1}) in the compact harmonic form

\be
\label{ZZ3}
\frac{d^2{\cal V}}{dx^2}+\left(\alpha+\frac{1}{2}-\frac{1}{4}x^2\right){\cal V}=0
\ee
The normalized square integrable solutions to (\ref{ZZ3}) are parabolic cylinder functions $D_\alpha(x)$
(solutions to the harmonic problem)

\be
\label{ZZ4}
f_1(\alpha, x)=\frac{(2\omega_0\tilde\nu)^{\frac{1}{4}}\sqrt{2}}{(4f_0\kappa\Gamma(\alpha+1)\sqrt{2\pi})^{\frac{1}{2}}}D_{\alpha}(x)
\ee
In the heavy quark limit with $\tilde\nu\rightarrow 0$, the spectrum in (\ref{ZZ3}) is harmonic 
for $p^2=-m_n^2$, and identical to 
the harmonic spectrum in (\ref{22XX15}-\ref{22XX16}).

\subsubsection{Axial decay constants: $f_{2k+1}$}

For the odd harmonic states with $\alpha=2k+1$,  we have

\be
\left(\frac{df_1}{dz}\right)(\alpha, 0)=\frac{(2\omega_0\tilde \nu)^{\frac{3}{4}}\sqrt{2}(2k+1)!!}{(4f_0\kappa(2k+1)!\sqrt{2\pi})^{\frac{1}{2}}}(-1)^{k}\nonumber\\
\ee
which gives through the Wronskian

\be
f_2(\alpha, 0)=\frac{1}{2\kappa g_0}\frac{(4f_0\kappa(2k+1)!\sqrt{2\pi})^{\frac{1}{2}}}{\sqrt{2}(2k+1)!!(2\omega_0\tilde \nu)^{\frac{3}{4}}}(-1)^{k+1}
\ee
Also, in the vicinity of $\alpha=2k+1$ we note that, 

\be
\alpha-2k-1=-\frac{p^2+m_{2k+1}^2}{\omega_0\tilde \nu}\frac{f_0}{g_0}
\ee
as well as

\bea
f_1(\alpha, 0)= &&\frac{(2\omega_0\tilde\nu)^{\frac{1}{4}}\sqrt{2}(2k)!!}{(4f_0\kappa(2k+1)!\sqrt{2\pi})^{\frac{1}{2}}}
\nonumber\\&&\times \frac{\sqrt{2\pi}}{2}(-1)^{k+1}\times (\alpha-2k-1)
\eea
Combining the above results, we finally find that

\bea
\label{FFHL1}
&&f_{2,asy}^2(2k+1,z_\Lambda)f_{2k+1}^2=\nonumber \\ 
&&\frac{1}{2\kappa g_0}\frac{4f_0\kappa(2k+1)!\sqrt{2\pi}}{(2k+1)!(2\omega_0\tilde\nu)\sqrt{2\pi}}\frac{g_0(\omega_0\tilde \nu)}{f_0}=1
\eea

\subsubsection{Vector decay constants: $f_{2k}$}

A rerun of  the previous steps for the even harmonic states with $\alpha=2k$ gives

\be
f_1 (\alpha,0)=(2\omega_0\tilde\nu)^{\frac{1}{4}}\frac{\sqrt{2}(2k-1)!!}{(4f_0\kappa(2k)!\sqrt{2\pi})^{\frac{1}{2}}}(-1)^{k}
\ee
which combined with the Wronskian gives

\be
\left(\frac{df_2}{dz}\right)(\alpha, 0)=\frac{1}{2\kappa g_0}\frac{(4f_0\kappa(2k)!\sqrt{2\pi})^{\frac{1}{2}}}{\sqrt{2}(2k-1)!!(2\omega_0\tilde \nu)^{\frac{1}{4}}}(-1)^{k}
\ee
Also near $\alpha=2k$ we have

\bea
\frac{d}{dz}f_1(\alpha, 0)=(2\omega_0\tilde\nu)^{\frac{3}{4}}\frac{\sqrt{2}(2k)!!(-1)^{k}\sqrt{2\pi}}{(4f_0\kappa(2k)!\sqrt{2\pi})^{\frac{1}{2}}} (\alpha-2k-1)\nonumber\\
\eea
and therefore

\be
\label{FFHL2}
f_{2,asy}^2(2k,z_\Lambda)f_{2k}^2=1
\ee


\subsubsection{Estimate}

The dependence on the cutoff $z_\Lambda$ reflects on the dependence  on the
heavy quark mass  which is $M$ in the large mass limit. In our D-brane set up
in section II we have made the assumption that the stringy HL modes are
approximated by local bi-fundamental vector fields in the world-volume of the light branes. 
This approximation precludes us from a rigorous evaluation of  the dependence
of $z_\Lambda$.   Qualitatively, we may estimate $z_\Lambda(M)$ by identifying
$M$ with the mass of a straight Nambu-Goto string pending from the heavy 
$8_H$-brane to the light $8_L$-branes as shown in Fig.~\ref{fig_branex}. 
In the Einstein frame and for large $z_\Lambda$, we have

\bea
\label{10X66}
M\approx &&\frac 1{2\pi l_s^2}\int_0^{z_\Lambda}  dz \left(-g_{tt}g_{zz}\right)^{\frac 12}\nonumber\\
\approx &&\frac 1{2\pi l_s^2}\int_0^{z_\Lambda}  dz \left(\frac {4U_{KK}}{9U_z}\right)^{\frac 12}
\approx \frac{1}{\pi l_s^{4\over 3}}
\left(\frac 29\lambda M_{KK}z_\Lambda^2\right)^{\frac{1}{3}}\nonumber\\
\eea
which shows that $z_\Lambda\approx M^{\frac 32}$ in our estimate. 
Using (\ref{10X66})  and the explicit form of $f_2$ where the overall constant is fixed
by the Wronskian, we can make an estimate of the decay constants in (\ref{FFHL1}) and (\ref{FFHL2}).  
The vector (n-even) and axial-vector (n-odd) decay constants are

\bea
\label{FFHL3}
f_n(M)=\frac{f(n)}{\sqrt{n!}}\frac{2^{n-\frac{33}{32}}}{\pi^{37\over 16}3^{\frac{25}{18}}}
\frac{{\mathbb C}_M^{\frac{n}{2}+\frac {11}{16}}}{e^{\mathbb C_M}}\sqrt{N_c}\,\lambda^{17\over 16}\,
M_{KK}^2
\eea
Here $\mathbb C_M$ is the  dimensionless combination

\be
\label{FFHL33}
\mathbb C_M= \frac 1{\sqrt{2}}\left(\frac{9\pi }{\lambda}\right)^3\left(\frac{M}{2M_{KK}}\right)^4
\ee
with $M_{KK}=m_\rho/\sqrt{0.67}=1.22\,m_\rho=941$ MeV. The constant $f(n)$ is related to the expansion
of the parabolic cylinder functions. Note that  (\ref{FFHL3}) is of order $\sqrt{N_c}$.

The holographic result (\ref{FFHL3}-\ref{FFHL33}) holds in the heavy quark limit. In particular,  we note that
the ratio of the pseudo-scalar  D to B meson decay constants following from (\ref{FFHL3}-\ref{FFHL33}) 
using the canonical definition $f_{Q_n}=f_n/m_n$ and to  leading order in $1/\lambda$, is

\be
\label{RAT}
\frac{f_{B_n}}{f_{D_n}}=\left(\frac{m_B}{m_D}\right)^{2n+\frac {7}{4}}\left(1+{\cal O}\left(\frac 1{\lambda^3}\right)\right)
\ee
For $n=0$, (\ref{RAT}) is to be compared to   
$f_B/f_D=(m_D/m_B)^{\frac 12}$ from general arguments~\cite{ISGUR}. 
While the two ratios reduce to 1 in the heavy mass limit, they differ sharply at finite masses
owing to our crude estimate in  (\ref{10X66}), and more generally our use
of  local bi-fundamental fields  to describe non-local string modes. This last concern can be altogether
bypassed in the bottom-up approach~\cite{HOLOLIU2}.

 \vfil

\end{document}